\documentclass[showpacs,aps,prl,twocolumn,superscriptaddress]{revtex4}

\usepackage{graphicx}
\usepackage{amsmath, amssymb}

\begin{document}

\title{Dressed-atom multiphoton analysis of anomalous electromagnetically induced absorption}

\author{Hsiang-Shun \surname{Chou}}
\affiliation{Max-Planck-Institut f\"{u}r Kernphysik, Saupfercheckweg 1, 69117 Heidelberg, Germany}
\affiliation{Institute of Optoelectronic Sciences, National Taiwan Ocean University, Keelung, Taiwan 202, R. O. C.}

\author{J\"{o}rg \surname{Evers}}
\affiliation{Max-Planck-Institut f\"{u}r Kernphysik, Saupfercheckweg 1, 69117 Heidelberg, Germany}

\date{\today}

\begin{abstract}
A method to interpret probe spectra of driven degenerate atomic systems is discussed. The dressed-atom multiphoton spectroscopy (DAMS) is based on a dressing of the atomic system with the strong coupling field, followed by a perturbative treatment of the probe field interaction. As example, we apply the DAMS to provide a clear interpretation for anomalous electromagnetically induced absorption (EIA), which cannot be explained by spontaneous transfer of coherence. We show that anomalous EIA arises from quantum interference between competing two-photon transitions, and explain the different dependences on the coupling field strength observed for various angular momentum setups.
\end{abstract}


\maketitle


One of the primary tools to study atomic systems is the measurement of probe spectra. Im particular, coherently laser-driven atomic systems offer a multitude of fascinating features when studied with a weak probe field. Driven degenerate systems exhibit large nonlinearities and rich quantum interference effects. Therefore, theoretical methods which carefully account for the nonlinear response and quantum interference are required to analyze and explain the dynamics of driven degenerate systems. Among the most important coherence effects revealed by this type of spectroscopy are electromagnetically induced transparency (EIT)~\cite{1,n6,n7} and the related electromagnetically induced absorption (EIA)~\cite{n6,n7,2,3,n3,n2,n4,4,5,6,7,8,9,10,n1,n5,11,12}.  EIA is a quantum optical effect that leads to a substantial enhancement of the absorption rate of a probe field by the presence of a coupling field. In the following, we will use it as prominent example for an as-yet unexplained nonlinear feature of driven degenerate atomic systems, and will explain its origin using the main result of this work, a dressed-atom multiphoton approach to atomic spectroscopy.

Based on experimental results and numerical calculations, the authors of  Ref.~\cite{3} concluded that EIA occurs in a degenerate two-level system provided that three requirements are satisfied: (i) $F_{e}\!=\!F_{g}\!+1\!$ ($F_{g}$ and $F_{e}$ are the total angular momentum of the ground and 
excited levels, respectively); (ii) transition $g\rightarrow e$ must be closed; (iii) $F_{g}>0$.  Later, it was argued~\cite{8} that EIA is caused by the transfer of coherence (TOC) between the excited and ground levels via spontaneous decay.  Experimental evidence in support of the argument presented
in Ref.~\cite{8} is provided by demonstrating that EIA resonances are suppressed (and even reversed) if the coherence of the excited state is significantly destroyed by collisions before the occurrence of the spontaneous decay~\cite{9}. The experimental conditions for EIA were rationalized by Goren {\it et. al.}~\cite{10}, who showed that TOC can only take place for systems where the population is not trapped in the ground level, i.e., for systems with $F_{e}\!=\!F_{g}\!+\!1$.  But surprisingly, in a recent experiment, EIA violating conditions (i) and (ii) were observed in several transitions of the Rb $D_{1}$ line with $F_{e}\!=\!F_{g}-1$ and  $F_{e}\!=\!F_{g}$~\cite{11}. In the following, we call this type of EIA {\it anomalous EIA}.

Motivated by the lack of explanation of this experimental result, in this Letter, we propose a method to interpret probe absorption spectra based on a perturbative expansion in a suitable dressed state picture of the system. As example, we show that this dressed-atom multiphoton spectroscopy (DAMS)
provides a straightforward interpretation for the  anomalous  EIA in degenerate two-level systems. More generally, we are interested in the absorption spectrum of a weak probe field through atoms irradiated by an intense coupling field.  In the DAMS, the coupled system ``atoms+coupling photons'' is described by the dressed-state wave functions which describe the coupling field to all orders. The interaction between atoms and the probe field is treated in perturbation theory. We show that the different features of the probe spectrum can easily be attributed to different classes of multiphoton transitions between bare and dressed states. As an advantage of our approach, quantum interferences among competing multiphoton pathways naturally arise in DAMS, enabling one to explain the occurrence of features with sub-natural line widths.

\begin{figure}[t]

\includegraphics[width=8cm]{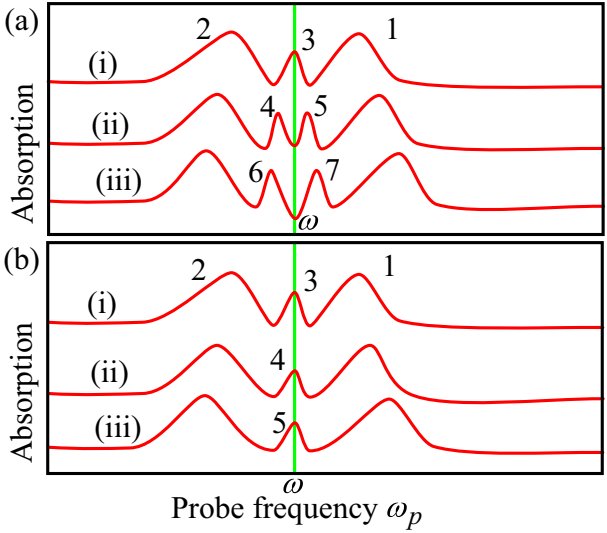}

\caption{\label{fig:t1}(Color online) A schematic depiction for the measured spectra exhibiting anomalous EIA in systems with (a) $F_{e}\!=\!F_{g}\!-\!1$ and (b) $ F_{e}\!=\!F_{g}\!=\!1$ (see FIGs. 4 and 5 of Ref.~\cite{11}). 
The curves show (i) low, (ii) intermediate, and (iii) high coupling field intensity. 
At low intensity, in both cases  an EIA peak is observed within an absorption dip. Increasing the intensity, the EIA peak breaks up in (a), but not in (b).  Here $\omega$ is the atomic frequency.}
\end{figure}

In the experiment of Kim {\it et. al.}~\cite{11,note}, the probe and the coupling beams were prepared in perpendicular linear polarizations, and the frequency of the coupling laser was locked to the atomic frequency $\omega$.  A schematic representation of anomalous EIA in systems with $F_{e}\!=\!F_{g}\!-\!1$ is depicted in Fig.~\ref{fig:t1}(a).  At  low coupling intensity, an EIA peak (peak 3 in Fig.~\ref{fig:t1}(a)) was observed within an absorption dip.  At intermediate coupling intensity, the EIA peak  broke up into two peaks (peaks 4 and 5 in Fig.~\ref{fig:t1}(a)) with a narrow dip between them. At high coupling intensity, the  dip became broader and deeper (curve (iii) in Fig.~\ref{fig:t1}(a)). 
A schematic figure for the anomalous EIA in systems with $F_{e}=F_{g}=1$ is depicted in Fig.~\ref{fig:t1}(b).  Here, an anomalous EIA peak was observed within an absorption dip.  
But in contrast to the previous case, the EIA peak did not break up into two peaks at intermediate and high coupling intensities.
Further measurements showed that in the transitions $F_{g}\!=\!F_{e}=2$  and $F_{g}\!=\!F_{e}\!=\!3$, the EIA peak broke up again at intermediate coupling intensity.

Density matrix calculations including TOC have been performed to first order in the probe field \cite{12}. The calculations describe successfully the {\it  normal}  EIA with $F_{e}\!=\!F_{g}\!+\!1$, but did not predict anomalous EIA with $F_{e}\!=\!F_{g}\!-\!1$ and 
$F_{e}=F_{g}$.  This suggests that anomalous EIA may be due to higher-order processes.

We now explain DAMS, and demonstrate that it  provides a straightforward interpretation for all observed features of anomalous  EIA. In the frame whose z axis is along the direction of the coupling polarization, the coupling field is $\pi$
polarized, whereas the probe polarization has $\sigma^{+}$ and $\sigma^{-}$ components.  In the first step, we describe the coupled system ``atoms+coupling photons'' by the dressed-state wave functions
\begin{subequations}
\label{dressed-state}
\begin{align}
|m_{F}+,n\rangle&=\frac{1}{\sqrt{2}}\left (e^{-i\phi}|m_{F},n+1\rangle +|m_{F^{\prime}},n\rangle \right)\,, 
\\
|m_{F}-,n\rangle&=\frac{1}{\sqrt{2}} \left(|m_{F},n+1\rangle -e^{i\phi}| m_{F^{\prime}},n\rangle \right)\,,
\end{align}
\end{subequations}
where $m_{F^\prime}=m_F$ and $n$ refers to the number of coupling photons present.  In Eqs.~(\ref{dressed-state}), $\phi$ is the phase of the dipole matrix element $g_{i}$ given by
\begin{equation}
g_{i}=-i\sqrt{\frac{\omega}{2\varepsilon_{0}\hbar V}}\langle i^{\prime}|
\vec{d}\cdot\hat{\varepsilon}_{c}|i\rangle=|g_{i}|\,e^{i\phi}\,.
\end{equation}
Here, $\vec{d}$ is the dipole moment operator and $\hat{\varepsilon}_{c}$ is the coupling field polarization. We use the abbreviation $|i\rangle\equiv|m_{F}=i\rangle$ and $|i^{\prime}\rangle\equiv|m_{F^{\prime}}=i\rangle$. The pair of dressed states are separated by an energy spacing $\hbar\Omega_{i}$, where $\Omega_{i}=2|g_{i}|\sqrt{n}$ is the Rabi frequency.  

%
%
\begin{figure}[t]
\centering

\includegraphics[width=8cm]{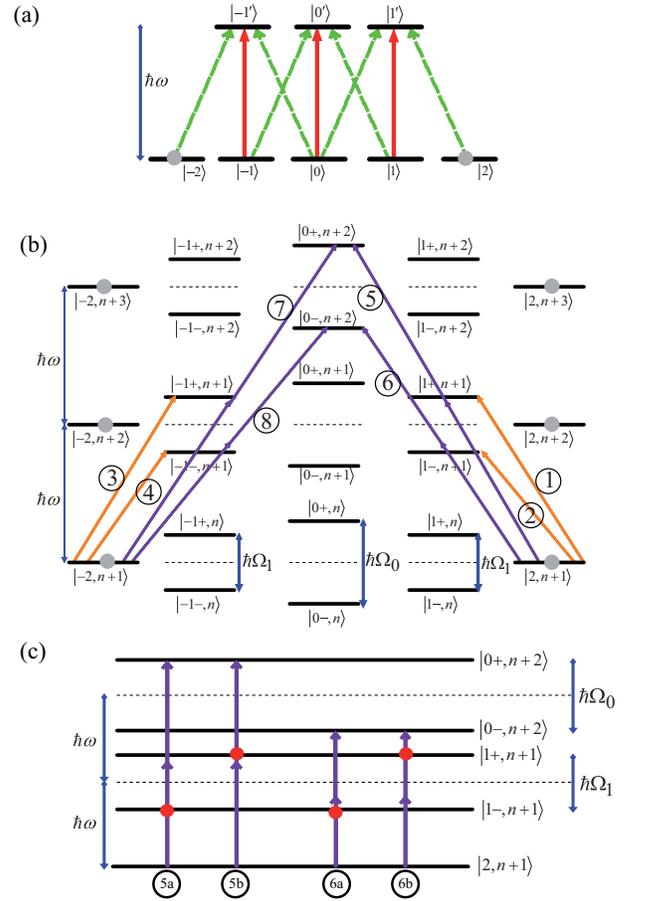}

\caption{\label{fig:t3}(Color online) Energy level diagram for systems with $F_{e}=F_{g}-1=1$. (a) Bare-atom picture. Solid red and dashed green arrows indicate the coupling and probe excitations, respectively. The solid circles indicate the population distributions without probe field; the atom is trapped in the states $|2\rangle$ and  $|-2\rangle$.
(b) Dressed-atom picture. The dotted lines indicate the unperturbed energy levels, solid circles indicate the population distributions without probe field. Encircled numbers label the relevant transitions induced by the probe field. Transitions 1-4 are one-photon bare-to-dressed transitions, 5-8 are two-photon bare-to-dressed transitions.
(c) Two-photon excitations between the bare state $|2,n+1\rangle$ and the dressed states $|0\pm,n+2\rangle$ .  Dotted lines indicate the unperturbed energy levels, and solid circles the intermediate states.}
\end{figure}
%

In the second step, we use perturbation theory to treat the probe field.
We start by discussing transitions with $F_{e}\!=\!F_{g}-1$.  As a typical example, we study the transition  $5 ^{2}S_{\frac{1}{2}} F_{g}\!=\!2\rightarrow 5 ^{2}P_{\frac{1}{2}} F_{e}\!=\!1$.  In Fig.~\ref{fig:t3}(a), we show the excitation diagram in the bare-atom picture. The solid circles indicate the population distributions without probe field.  In the present case, the populations are trapped
in the states $|2\rangle$ and  $|\!-2\rangle$.  In Fig.~\ref{fig:t3}(b), we show the excitation diagram in the dressed-atom picture. 
We find that the first class of relevant processes are one-photon transitions from a bare to a dressed state (1BD). These are the transitions
$|2,n+1\rangle  \rightarrow |1\pm,n+1\rangle$ and $|-2,n+1\rangle  \rightarrow |-1\pm,n+1\rangle$ indicated by arrows 1 to 4 in Fig.~\ref{fig:t3}(b). Due to the population distribution, these processes lead to two absorption peaks at $\omega_{p}=\omega\pm\Omega_{1}/2$, which are clearly visible as peaks 1 and 2 in Fig.~\ref{fig:t1}(a). 
All other processes are of higher order, and the next relevant class are two-photon bare-to-dressed transitions (2BD). These are the transitions $|2,n+1\rangle  \rightarrow |0\pm,n+2\rangle$ and
$|-2,n+1\rangle  \rightarrow |0\pm,n+2\rangle$ shown as arrows 5 to 8 in Fig.~\ref{fig:t3}(b). The two 2BD transitions 5 and 6 are shown in more detail in Fig.~\ref{fig:t3}(c). There are two competing transitions, which differ in the intermediate states, between a given initial and final states.  The relevant intermediate states are  indicated by the circles in Fig.~\ref{fig:t3}(c). For example, transitions 6a and 6b at 
$\omega_{p}=\omega-\Omega_{0}/4$ take place between $|2,n+1\rangle$ and $|0-,n+2\rangle$.
The intermediate states for these transitions are $|1-,n+1\rangle$ and  $|1+,n+1\rangle$, respectively.  Their transition amplitudes evaluate to
\begin{subequations}
\begin{align}
T^{(1)}&=\frac{\langle 0-,n+2|V|1-,n+1\rangle\langle 1-,n+1|V|2,n+1\rangle}
{E(1-,n+1)-E(2,n+1)-\hbar(\omega-\frac{\Omega_{0}}{4})}\nonumber\\
&=\frac{\sqrt{2}\exp (-2i\phi)\langle 0^{\prime}|V|1\rangle\langle 1^{\prime}|V|2
\rangle}{-\hbar(2\Omega_{1}-\Omega_{0})} \,,
\\
T^{(2)}&=\frac{\langle 0-,n+2|V|1+,n+1\rangle\langle 1+,n+1|V|2,n+1\rangle}
{E(1+,n+1)-E(2,n+1)-\hbar(\omega-\frac{\Omega_{0}}{4})}\nonumber\\
&=\frac{-\sqrt{2}\exp (-2i\phi)\langle 0^{\prime}|V|1\rangle\langle 1^{\prime}|V|2
\rangle}{\hbar(2\Omega_{1}+\Omega_{0})} \,,
\end{align}
\end{subequations}
where $V$ is the interaction Hamiltonian due to the probe field.  The two-photon absorption rates are substantially enhanced since the probe field is nearly resonant with the intermediate states.  Moreover, it is interesting to see that $T^{(1)}$ and $T^{(2)}$ have the same phase, which means that  constructive interference takes place. A similar analysis shows that constructive interference
also occurs between transitions 5a and 5b at $\omega_{p}=\omega+\Omega_{0}/4$.
The resonant enhancement together with the constructive quantum interference increase the two-photon absorption rates substantially, such that the two-photon absorptions become relevant in the driven degenerate atomic systems as the EIA peaks 3-7 in Fig.~\ref{fig:t1}(a). 
Finally, we analyze the dependence of the EIA peak on the coupling intensity. At low coupling intensity ($\Omega_0 < \Gamma$), the two absorption peaks overlap at the line center and produce a two-photon EIA peak (peak 3 in Fig.~\ref{fig:t1}(a)) within the one-photon absorption dip.   
The separation of the two peaks increases with the coupling intensity.  At intermediate coupling intensity ($\Omega_0 \approx \Gamma$), the EIA peak breaks up into two peaks (peaks 4 and 5 in Fig.~\ref{fig:t1}(a)) with a narrow dip between them.  At high coupling intensity ($\Omega_0 > \Gamma$), the dip broadens and deepens (curve (iii) in Fig.~\ref{fig:t1}(a)).  
We predict 
that the EIA peak breaks up into four absorption peaks as the coupling intensity increases,
in the case that the polarizations of the coupling and 
the probe fields are not perpendicular.  This is due to the fact that the
2BD transitions $|2,n+1\rangle\rightarrow |1\pm,n+2\rangle$ and 
$|-2,n+1\rangle\rightarrow |-1\pm,n+2\rangle$, which are induced by the $\pi$
component of the probe field, give rise to additional absorption peaks at 
$\omega_{p}=\omega\pm\Omega_{1}/4$. 
%

\begin{figure}[t]
\centering

\includegraphics[width=8cm]{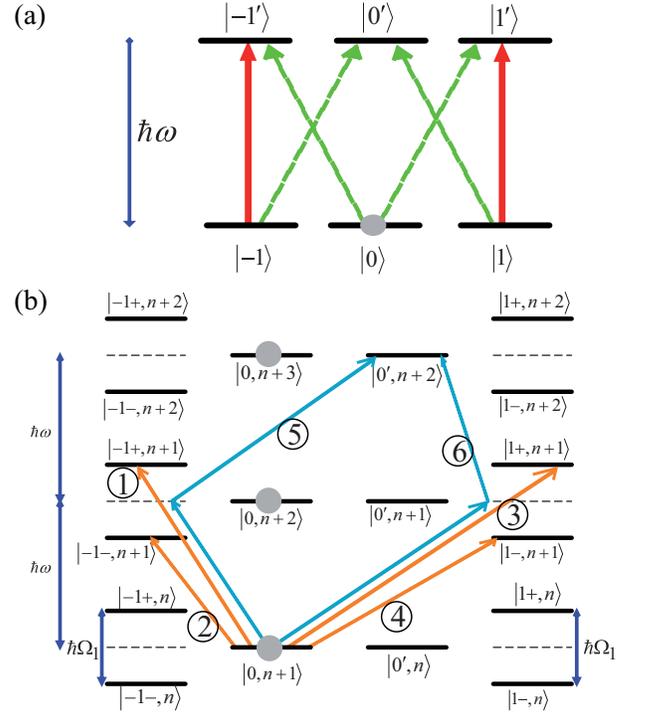}

\caption{\label{fig:t6}(Color online) 
(a) Bare-atom level scheme for systems with $F_{e}\!=\!F_{g}\!=\!1$.  
Solid red and dashed green arrows indicate the coupling and probe
excitations, respectively.  The transition $|0\rangle\rightarrow |0^{\prime}\rangle$ is forbidden. The solid circle indicates the population distribution without probe field; the population is trapped in $|0\rangle$.
(b) The corresponding scheme in the dressed-atom picture.  The dotted lines indicate the unperturbed energy levels.  The arrows indicate the relevant probe field excitations. Channels 1-4 are one-photon bare-to-dressed transitions, while channels 5 and 6  are two-photon bare-to-bare transitions.}
\end{figure}

We now proceed with our analysis in transitions with  $F_{e}=F_{g}$.  First, we study the transition $5 ^{2}S_{\frac{1}{2}} F_{g}=1\rightarrow 5 ^{2}P_{\frac{1}{2}} F_{e}=1$.  The excitation diagram in the bare-atom picture is shown in Fig.~\ref{fig:t6}(a).  The transition $|0\rangle\rightarrow |0^{\prime}\rangle$ is forbidden.  As a result, the population is trapped in $|0\rangle$.  The excitation diagram in the dressed-atom picture is shown in Fig.~\ref{fig:t6}(b).  
Again, the leading order processes are one-photon bare-dressed (1BD) transitions, namely
$ |0,n+1 \rangle \rightarrow |-1\pm,n+1\rangle$ and $|0,n+1\rangle\rightarrow |1\pm,n+1\rangle$,
indicated as arrows 1-4 in Fig.~\ref{fig:t6}(b). These transitions lead to two absorption peaks at $\omega_{p}=\omega \pm \Omega_{1}/2$, shown as peaks 1 and 2 in Fig.~\ref{fig:t1}(b).
But in contrast to the previous case, it turns that the anomalous EIA can be attributed to two-photon transitions between bare states (2BB) $|0,n+1\rangle\rightarrow|0^{\prime},n+2\rangle$ at $\omega_{p}=\omega$, indicated as arrows 5 and 6 in Fig.~\ref{fig:t6}(b). There are four competing 2BB transitions, which differ in the intermediate states  $|-1-,n+1\rangle$, $|-1+,n+1\rangle$, $|1-,n+1\rangle$, or $|1+,n+1\rangle$.  It is straightforward to show that the competing transitions have the same transition amplitude. The phases of the transition amplitudes indicate that constructive interference enhances the two-photon absorption rate, giving rise to a two-photon EIA peak shown as peak 3 in Fig.~\ref{fig:t1}(b). 

The crucial difference to the previous case is now that there is only one absorption frequency in the 2BB processes, since the bare states are unaffected by the applied coupling field. Even at high coupling intensity, the EIA peak does not break up into two peaks as it did in the previous case. Thus, the DAMS naturally explains the difference between anomalous EIA in degenerate two-level systems with different  angular momentum setups.

\begin{figure}[t]
\includegraphics[width=0.9\columnwidth]{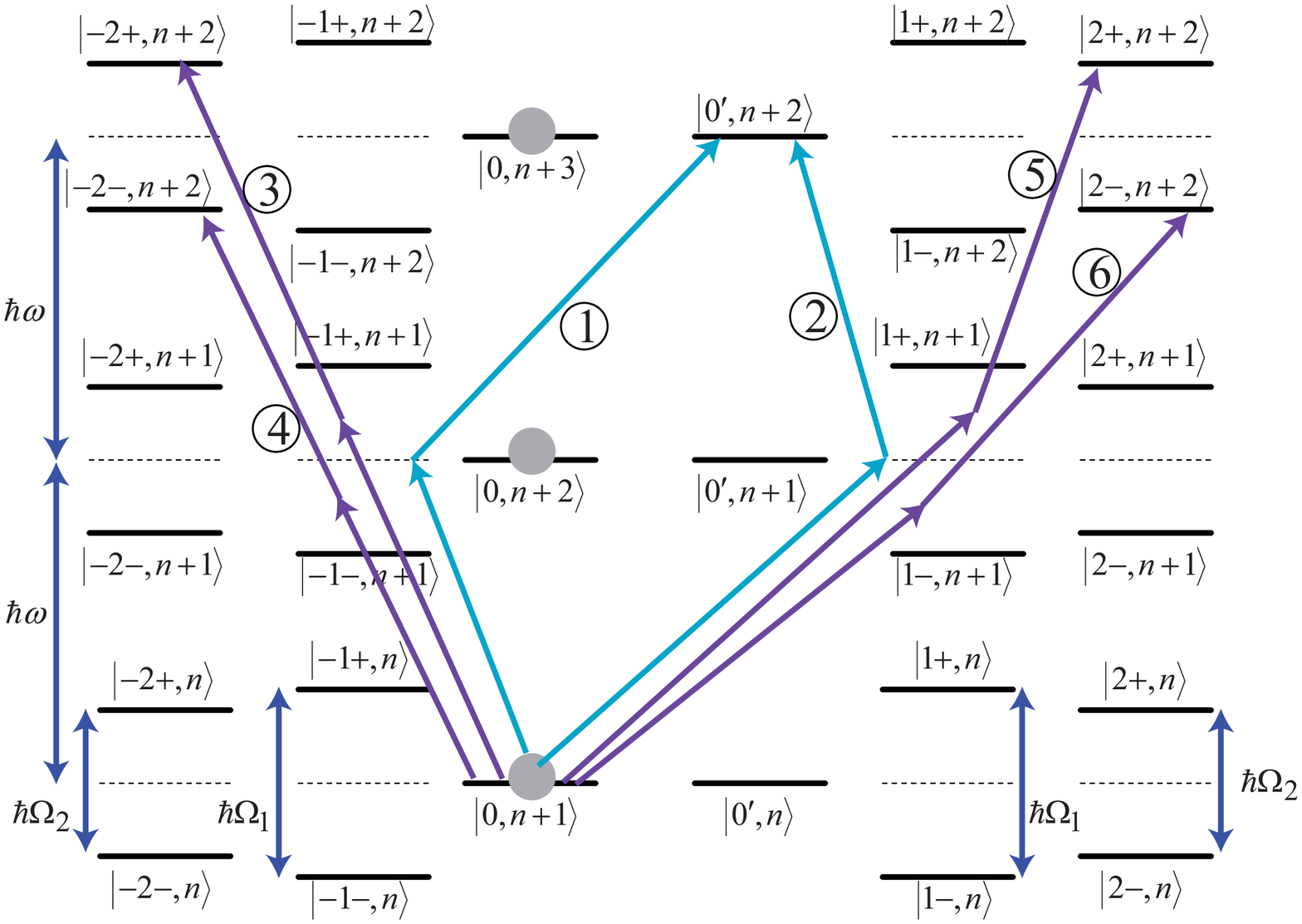}

\caption{\label{fig:t8}(Color online) Dressed-atom representation for systems with 
$F_{e}=F_{g}=2$. The dotted lines indicate the unperturbed energy levels.  The solid circles indicate the population distribution without probe field. The arrows indicate the relevant multiphoton transitions. Channels 1 and 2 are two-photon bare-to-bare transitions, and channels 3-6 are two-photon bare-to-dressed transitions. The one-photon transitions are omitted for clarity.}
\end{figure}

We continue to explore the transition $5 ^{2}S_{\frac{1}{2}} F_{g}=2\rightarrow 5 ^{2}P_{\frac{1}{2}} F_{e}=2$. The excitation diagram in the dressed-atom picture is shown in Fig.~\ref{fig:t8}. The discussions follow similar lines as those for the systems with $F_{e}=F_{g}=1$. Extra absorption peaks, which are due to two-photon bare-to-dressed (2BD) transitions $|0,n+1\rangle \rightarrow |-2\pm,n+2\rangle$ and 
$|0,n+1\rangle \rightarrow |2\pm,n+2\rangle$ shown as arrows 3 to 6 in Fig.~\ref{fig:t8}, occur at $\omega_{p}=\omega\pm\Omega_{2}/4$.  In contrast to the two previous cases, these transitions occur at three different frequencies, namely $\omega_{p}=\omega$ and $\omega_{p}=\omega\pm\Omega_{2}/4$. At low coupling intensity, these peaks nearly overlap, and gives rise to a single two-photon EIA peak.  As the coupling intensity increases, the EIA peak breaks up into three peaks with narrow dips among them.  At high coupling intensity, the dips become broader and deeper.
We predict that the EIA peak breaks up into three and five absorption peaks for the systems
with $F_{e}=F_{g}=1$  and $F_{e}=F_{g}=2$ respectively, in the case that the polarizations
of the coupling and the probe fields are not perpendicular.  This is due to the fact that the
 two-photon BD transitions $|0,n+1\rangle\rightarrow |1\pm,n+2\rangle$
and $|0,n+1\rangle\rightarrow|-1\pm,n+2\rangle$ give rise to additional absorption
peaks at $\omega_{p}=\omega\pm \Omega_{1}/4$. 

In summary, we have discussed a method to interpret probe absorption spectra, which we term dressed-atom multiphoton spectroscopy. It is based on a representation of the atomic system in a dressed-atom picture with respect to the coupling field, followed by a perturbative analysis of the probe field.  Using DAMS, the different multiphoton transitions can be classified into n-photon bare-to-bare and bare-to-dressed state transitions. As example, we have applied DAMS to the example of anomalous EIA. We have demonstrated that our method provides a straightforward interpretation of anomalous  EIA, explaining the various features observed in transitions with different angular momentum configurations. The strong absorption follows from constructive interference among competing multiphoton transitions. The DAMS can readily be applied to shed new light on higher-order processes in other probe spectroscopy setups.

The authors thank Professors R. L. Chang,
J. B.  Kim and I. A. Yu for useful discussions. 




\begin{thebibliography}{99}
\bibitem{note}Note that the curve labeling of Figs.~4-6 in \cite{11} is reversed.
\end{thebibliography}
\end{document}